\documentclass[prx,twocolumn,showpacs,amsmath,amssymb,superscriptaddress,longbibliography]{revtex4-1}
\usepackage{psfrag,graphicx}
\usepackage{amsfonts,amssymb,amsmath}      
\usepackage{graphicx}
\usepackage{dcolumn}
\usepackage{bm}
\usepackage{float}
\usepackage{hyperref}
\usepackage{accents}
\usepackage{bbding}
\usepackage{color,soul}
\usepackage{epstopdf}
\usepackage{changepage}
\usepackage{orcidlink}
\usepackage[normalem]{ulem}

\newcommand{\erf}[1]{Eq.~(\ref{#1})}
\newcommand{\beq}{\begin{equation}}
\newcommand{\eeq}{\end{equation}}

\newcommand{\dg}{^\dagger}

\newcommand{\smallfrac}[2]{\mbox{$\frac{#1}{#2}$}}

\newcommand{\half}{\smallfrac{1}{2}}
\newcommand{\bra}[1]{\langle{#1}|}
\newcommand{\ket}[1]{|{#1}\rangle}
\newcommand{\ip}[2]{\langle{#1}|{#2}\rangle}
\newcommand{\op}[2]{\ket{#1}\bra{#2}}
\newcommand{\sch}{Schr\"odinger}

\newcommand{\Tr}{\text{Tr}}

\newcommand{\tp}{^{\top}}

\newcommand{\ex}[1]{\langle{#1}\rangle}
\newcommand{\dd}{{\rm d}}

\newcommand{\ie}{{\em i.e.}}

\newcommand{\ea}{{\em et al.}}

\newcommand{\past}[1]{\overleftarrow{#1}}
\newcommand{\fut}[1]{\overrightarrow{#1}}
\newcommand{\both}[1]{\overleftrightarrow{#1}}
\newcommand{\fil}{_{\text F}}
\newcommand{\rfil}{_{\text R}}
\newcommand{\sm}{_{\text S}}
\newcommand{\swv}{_{\rm SWV}}

\newcommand{\bx}{{\bf x}}

\definecolor{nblue}{rgb}{0.06,0.3,0.73}
\definecolor{nblack}{rgb}{0,0,0}
\definecolor{nred}{rgb}{0.9,0.1,0.1}
\definecolor{nmagenta}{rgb}{0.7,0.0,0.3}

\newcommand{\red}{\color{nred}}

\newcommand{\blk}{\color{nblack}}

\definecolor{applegreen}{rgb}{0.55, 0.71, 0.0}

\newcommand\stPW{\bgroup\markoverwith{\applegreen{\rule[0.5ex]{2pt}{0.4pt}}}\ULon}

\begin{document}

\title{Inequivalent ways to apply semi-classical smoothing to a quantum system}

\author{Kiarn T. Laverick\orcidlink{0000-0002-3688-1159}}
\affiliation{MajuLab, CNRS-UCA-SU-NUS-NTU International Joint Research Laboratory}
\affiliation{Centre for Quantum Technologies, National University of Singapore, 117543 Singapore, Singapore}

\author{Areeya Chantasri\orcidlink{0000-0001-9775-536X}}
\affiliation{Optical and Quantum Physics Laboratory, Department of Physics,
Faculty of Science, Mahidol University, 
Bangkok 10140, Thailand}
\author{Howard M. Wiseman\orcidlink{0000-0001-6815-854X}}
\affiliation{Centre for Quantum Computation and Communication Technology 
(Australian Research Council), \\ Quantum and Advanced Technologies Research Institute, Griffith 
University, Yuggera Country, Brisbane, Queensland 4111, Australia}

\date{\today}
\begin{abstract}

In this paper, we correct a mistake we made in [Phys.~Rev.~Lett.~{\bf 122}, 190402 (2019)] and [Phys.~Rev.~A~{\bf 103}, 012213 (2021)] regarding the Wigner function of the so-called smoothed Weak-Valued state (SWV state).
Here smoothing refers to estimation of properties at time $t$ using information obtained in measurements both before and after $t$. 
The SWV state is a pseudo-state (Hermitian but not necessarily positive) that gives, by the usual trace formula, the correct value for a weak measurement preformed at time $t$, \ie, its weak value. The Wigner function is a pseudo-probability-distribution (real but not necessarily positive) over phase-space. A smoothed (in this estimation sense) Wigner distribution at time $t$ can also be defined by applying classical smoothing for probability-distributions to the Wigner functions. The smoothed Wigner distribution (SWD) gives identical means for the canonical phase-space variables as does the SWV state. However, contrary to the assumption in the above references, 
the Wigner function of the SWV state is not the smoothed Wigner distribution.
\end{abstract}
\maketitle

\section{Introduction}
In recent years, the notion of applying the classical estimation technique of smoothing to quantum systems has been of great interest \cite{BaoMol20a,BaoMol20b,CGLW21,Tsa-PRA22,LWCW23}. Classical smoothing is an estimation technique that applies to dynamical open systems that are under continuous-in-time (weak) observation. Specifically, it estimates the state of the system at any time of interest, $t$, using measurement outcomes obtained both prior to and posterior to $t$ (the {\em past-future} record $\both{\bf O}$). In classical systems, the smoothed state can be computed by use of Bayes' theorem, where the state is proportional to the product of two different quantities, the filtered state (conditioned on the past measurement record $\past{\bf O}_t$) $\wp\fil(\bx;t):=\wp(\bx;t|\past{\bf O}_t)$ and the retrofiltered effect (the probability of the future record $\fut{\bf O}_t$) $E\rfil(\bx;t):=\wp(\fut{\bf O}_t|\bx;t)$. That is, 
\beq \label{classsm}
\wp\sm(\bx;t):= \wp(\bx;t|\both{\bf O}) = \frac{E\rfil(\bx;t)\wp\fil(\bx;t)}{\int d{\bf x} E\rfil(\bx;t)\wp\fil(\bx;t)}\,,
\eeq
where the denominator, from Bayes' rule, ensures normalization of the state.
In quantum systems, however, the concept of smoothing  is not so straightforward. 

{\blk One, semi-classical, approach to the} smoothing formula (\ref{classsm}) is the symmetric product of the quantum filtered state $\rho\fil(t)$ and the retrofiltered effect $\hat{E}\rfil(t)$, which are the analogues of the classical distributions \cite{Bel87,Bel99,Tsa-PRA09,GJM13}. That is, {where the subscript will be explained shortly,}
\beq\label{eq:swv_gen}
\varrho\swv(t) = \left(\hat{E}\rfil(t)\circ\rho\fil(t)\right)/N\,.
\eeq
Here, $\hat{A}\circ\hat{B}$ denotes the symmetric or Jordan product, namely  $\half (\hat{A}\hat{B}+\hat{B}\hat{A})$, and the normalization constant is 
\beq \label{Normsame}
N = {\rm Tr}\left[\hat{E}\rfil(t)\rho\fil(t)\right]\,.
\eeq
Equation (\ref{eq:swv_gen}) generally leads to a pseudo-state, \ie, a state that is not positive semidefinite. Nevertheless, 
this pseudo-state does have physical significance: taking the expectation value of {\em any} observable $\hat{A} = \hat{A}\dg$ with respect to $\varrho\swv(t)$ yields its weak-value (strictly the real part) at time $t$ \cite{GJM13,CGLW21}. That is, it correctly gives the expected value of an arbitrarily weak and minimally disturbing measurement of $\hat A$ performed at time $t$, conditioned on the prior and posterior measurements as encoded in $\rho\fil(t)$ and the retrofiltered effect $\hat{E}\rfil(t)$ respectively. 

The preceding fact is why we called the expression in \erf{eq:swv_gen} the Smoothed Weak-Valued (SWV) state,  as indicated by its subscript, while we used the variant Greek letter $\varrho$ (rather than $\rho$ as for $\rho\fil$) to indicate that it is a pseudo-state, when we introduced this notation in 
Ref.~\cite{LCW-PRA21}. However, the first consideration of such a pseudo-state can be traced back to Tsang in Eq.~(5.5) of Ref.~\cite{Tsa-PRA09}. 
There he considers the expectation of an arbitrary observable (as we have called $\hat A$) with the non-symmetrized product, $\hat E\rfil \rho\fil$ and identifies that with the (complex) weak-value of $\hat A$. Tsang's non-Hermitian pseudo-state subsequently appeared explicitly in the Supplementary Material of Ref.~\cite{GJM13}, where Gammelmark \ea\  referred to as the ``past density matrix".

It is also possible to define a smoothed quantum state that {\em is} positive semi-definite, and also has an operational meaning in terms of estimating unknown  measurement results, albeit one that is more subtle to explain \cite{LGW-PRA21,CGLW21,LWCW23}. Introduced by one of us and Guevara, we laid claim to the term ``quantum state smoothing'' for this procedure~\cite{GueWis15}, and $\rho\sm$ for the state. A different concept, which can be adapted to some of the same smoothing scenarios~\cite{CGLW21}, is the most likely path in quantum state space, introduced by one of us and co-workers \cite{CDJ13,CDJ15}.   
Finally, yet another definition for a valid quantum state that makes use of prior and posterior information, and which, like \erf{eq:swv_gen}, is defined only in terms of $\rho\fil$ and $\hat{E}\rfil$, has also been proposed recently by one of us and co-workers \cite{LASL25}. 

In this paper we are not concerned with the developments of the preceding paragraph except to note that the quantum state smoothing theory was applied by us to linear Gaussian quantum (LGQ) systems \cite{LCW19,LCW-PRA21}, and in this context, errors were made regarding $\varrho\swv$. Specifically, it was wrongly identified with yet another type of {\blk semi-classically} smoothed pseudo-state, based on smoothing the Wigner distribution. We discuss the smoothed Wigner distribution, as first introduced by Tsang~\cite{Tsa-PRA09}, and why it was analyzed in Refs.~\cite{LCW19,LCW-PRA21}, in Sec.~\ref{sec:SWD}. Then, in Sec.~\ref{sec:comp}, we compare its statistics with those of the smoothed weak-valued state, explain the error made in Refs.~\cite{LCW19,LCW-PRA21}, and illustrate the difference. 

\section{The Smoothed Wigner Distribution} \label{sec:SWD}

LGQ systems are quantum systems with linear dynamics and Gaussian noise in the Heisenberg picture, including in the measurement record. Such a statement only makes sense for systems in which there is a $2n-$vector $\hat{\bf x} = (\hat q_1,\hat p_1,\hat q_2,\hat p_2,\cdots,\hat q_n,\hat p_n)$, where $[\hat q_j, \hat p_k]=i\hbar\delta_{jk}$, forming a complete set of observables (in the estimation and control theory sense~\cite{WisMil10}). 
The quantum states (and suitably normalised effects) of LGQ systems are Gaussian~\cite{WisDoh05,ZhaMol17}, being described by only the mean and covariance of the quadrature operators $\hat{\bx}$ (to use a quantum optics expression). This simplicity allowed us to obtain closed-form expressions of the smoothed quantum state~\cite{LCW19,LCW-PRA21}. 

A natural way to define Gaussian states is through their Wigner function $W({\bf x})$, as this has the form of a Gaussian probability distribution in phase space, with the same mean and covariance. The general expression for the Wigner function for a state $\rho$ (not necessarily Gaussian) is 
\beq\label{eq:Wig_Cha}
W_\rho({\bf x}) = \frac{1}{(2\pi)^{2n}}\int \dd{\bf k}\,\,\chi_\rho({\bf k}) e^{-i{\bf k}\tp{\bx}}\,,\\
\eeq
where 
\beq
\chi_\rho({\bf k}) = \Tr\left[\rho e^{i{\bf k}\tp\hat\bx}\right]
\eeq
is the characteristic function (the Fourier transform of the Wigner function). In general \erf{eq:Wig_Cha} is a pseudo-probability-distribution, integrating to $\Tr[\rho]$ but not everywhere positive. But for Gaussian states it is everywhere positive, which invites a comparison with classical estimation theory. In particular, if one simply removes the hats from the Heisenberg equations for the system dynamics and measurements of a LGQ system, one obtains a classical theory for random phase-space variables with distributions equal to the Wigner functions. Applying the classical theory of smoothing (\ref{classsm}) would give a smoothed distribution 
\beq \label{SWD}
\digamma_{\rm SWD}(\bx;t) =  (2\pi)^n W\fil(\bx;t)W\rfil(\bx;t)/N\,,
\eeq
which is also Gaussian. Here $W\fil(\bx;t)$ and  $W\rfil(\bx;t)$ are short for $W_{\rho\fil}(\bx;t)$ and $W_{\hat E\rfil}(\bx;t)$ respectively, and the normalisation constant is identical to that in \erf{Normsame}.

Equation~(\ref{SWD}) was first proposed (with different notation) by Tsang \cite{Tsa-PRA09}, for arbitrary (not just Gaussian) Wigner functions, as a ``smoothing quasiprobability distribution''. Here we use the subscript SWD to stand for the more specific designation ``Smoothed Wigner Distribution''. But note that, similarly to \erf{eq:swv_gen}, we also use a different symbol, $\digamma$ (the archaic Greek letter wau), rather than $W$, for the normalised product. That is because $\digamma_{\rm SWD}(\bx;t)$ is not, in general, the Wigner function for any state.  To be specific, the Wigner-Weyl transform \cite{LamWil18} defines the pseudo-state 
\beq
\varrho_{\rm SWD}(t) = \frac{1}{(2\pi)^{n}} \int\dd{\bf k} \,\,\chi_{\rm SWD}({\bf k};t)e^{-i{\bf k}\tp\hat{\bx}}\,,
\eeq
where $\chi_{\rm SWD}({\bf k};t)$ is the Fourier transform of $\digamma_{\rm SWD}(\bx;t)$. This $\varrho_{\rm SWD}(t)$ is, in general, not positive-definite. This is so even if for LGQ systems where $\digamma_{\rm SWD}(\bx;t)$ is a positive Gaussian function; in general it is too concentrated in phase-space to correspond to a valid  state. Loosely, it violates the Heisenberg uncertainty relation \cite{Tsa-PRA09,LCW-PRA21}. 
Nevertheless, this classically inspired calculation gives an interesting comparison with quantum state smoothing for LGQ systems, where $\rho\sm$ is always a valid Gaussian state, and so we presented this comparison in Refs.~\cite{LCW19,LCW-PRA21}. 

\begin{figure*}[t]
\begin{minipage}{0.5\textwidth}
\includegraphics[scale = 0.36]{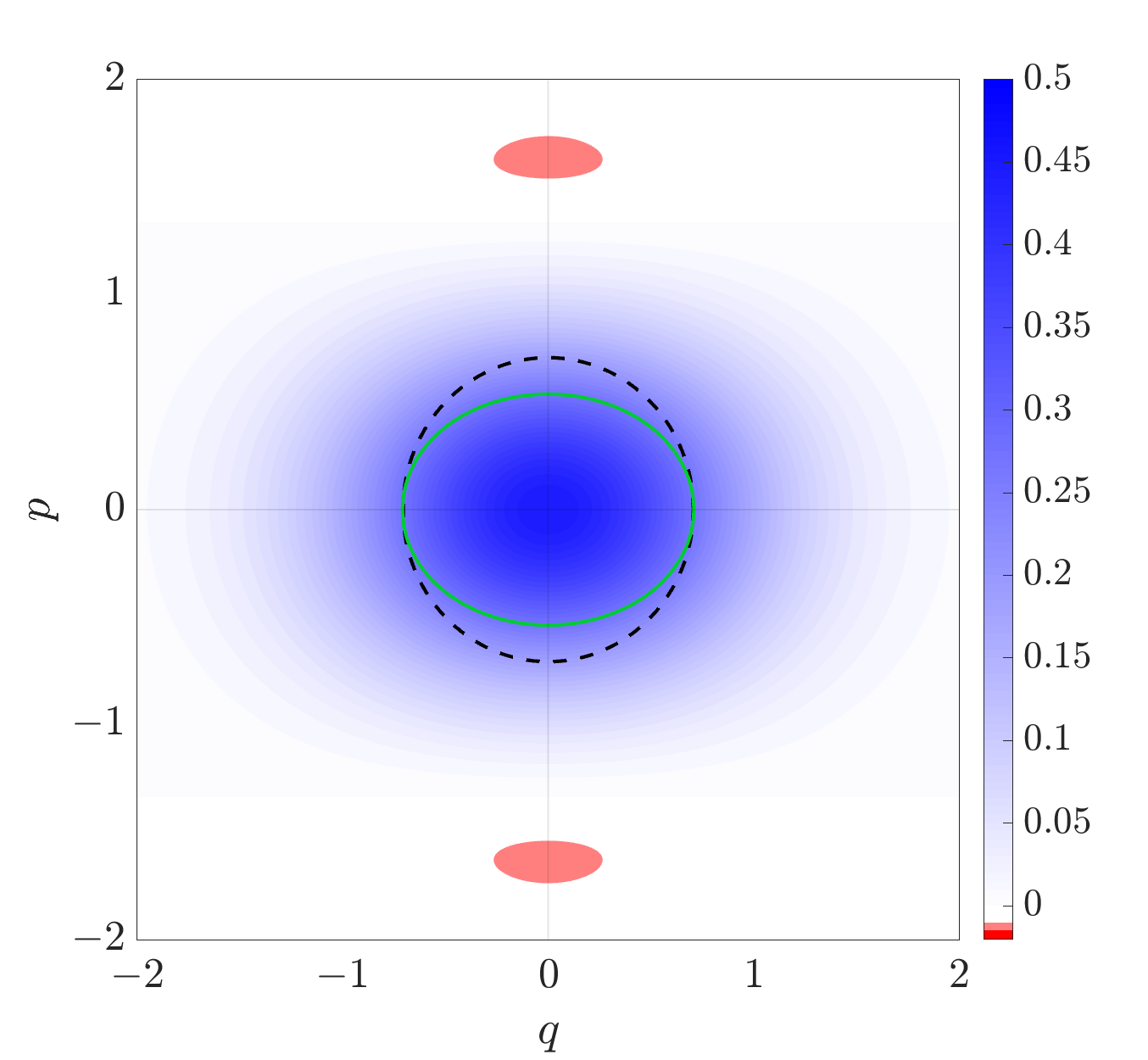}
\end{minipage}%
\begin{minipage}{0.5\textwidth}
\includegraphics[scale = 0.36]{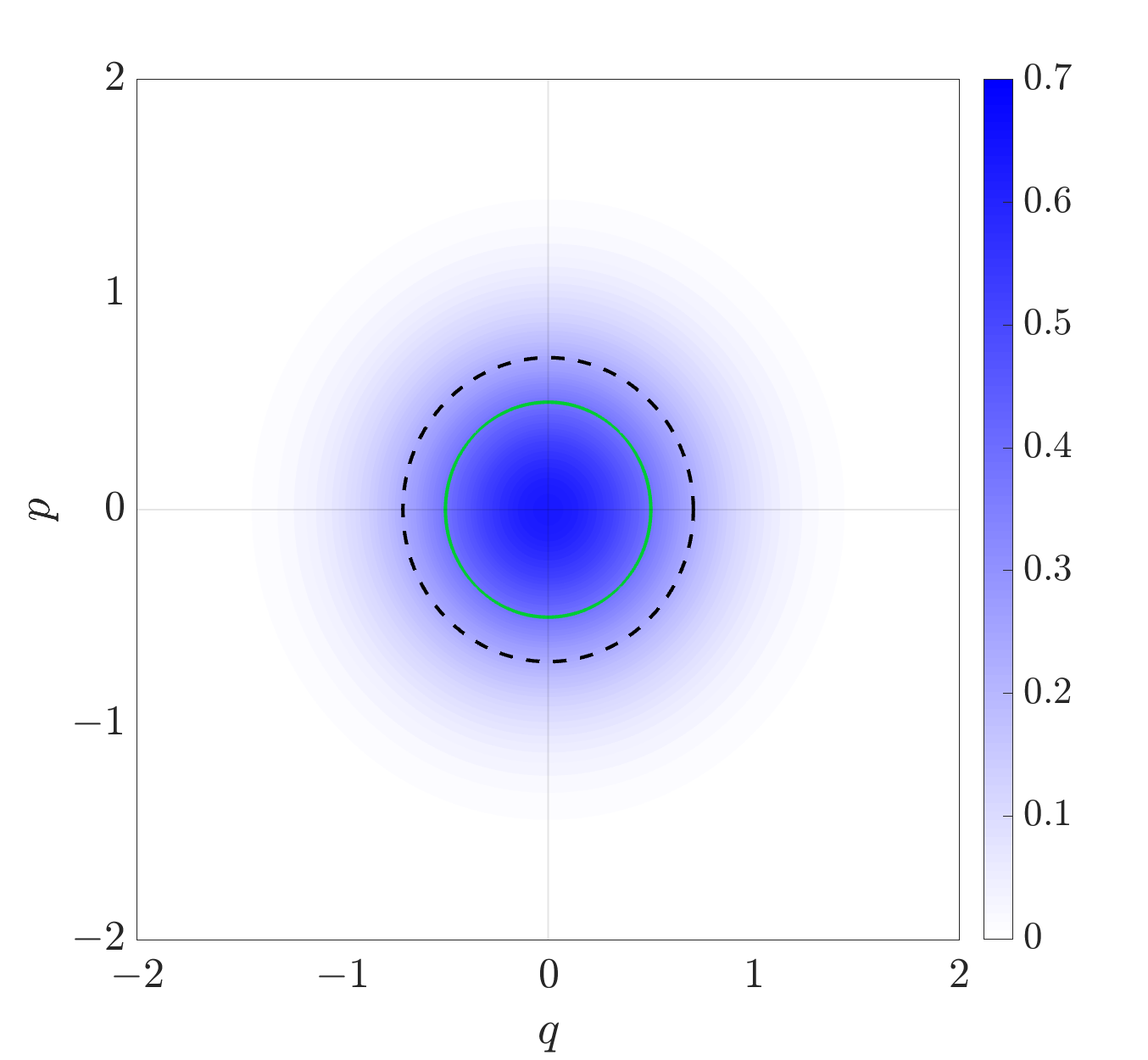}
\end{minipage}
\caption{Illustrative example of the Wigner function for the smoothed weak-value state (left) and the smoothed Wigner distribution (right) where  $\rho\fil = \ket{\alpha}\bra{\alpha}$ and $\hat{E}\rfil = \ket{-\alpha}\bra{-\alpha}$. In both plots the green curve is the $e^{-1/2}$-contour, 
which can be compared to the corresponding 
contour of the vacuum state (black dashed line). Here $\hbar = 1$ and we have taken $\alpha = \sqrt{\ln 2}/2$, which is a typical value, as explained in the text. }
\label{fig:cat_stripes}
\end{figure*}

\begin{figure}[t]
\includegraphics[scale = 0.36]{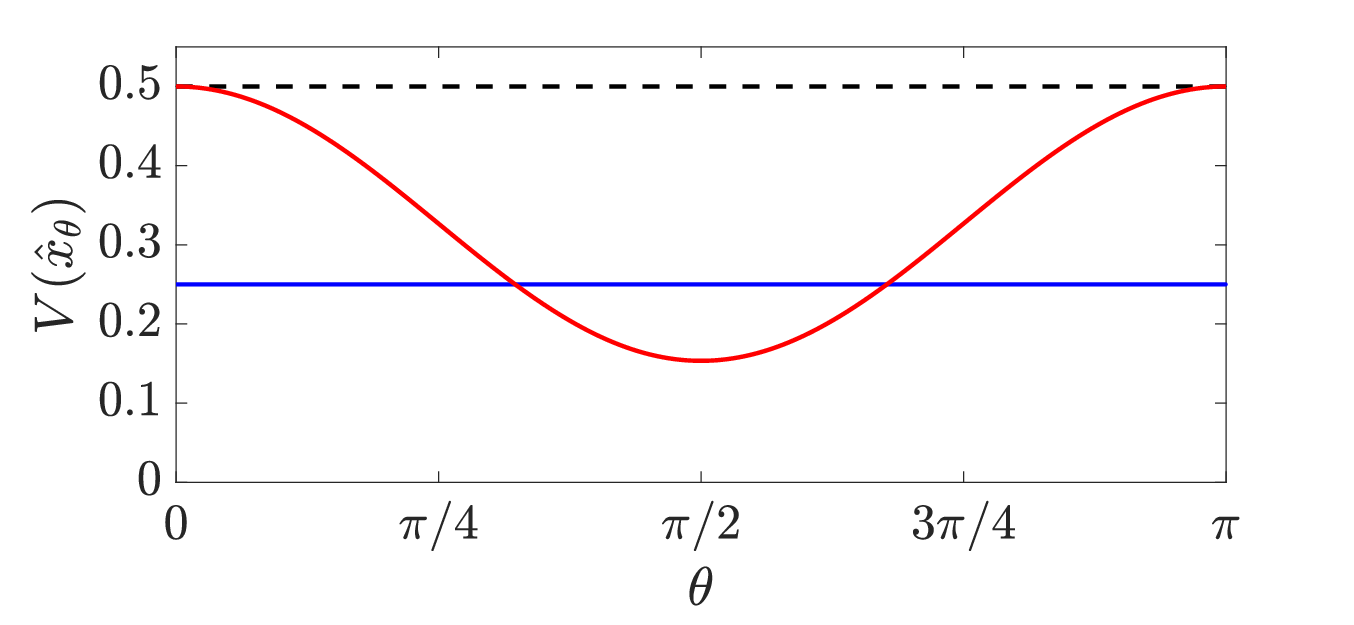}
\caption{The variance of the $\hat{x}_\theta = \hat{q}\cos\theta + \hat{p}\sin\theta$ quadrature for the SWD (blue line), SWV state (red line) and the vacuum state (black dashed line).
 Notably, the variance for the SWD (SWV state) is always smaller (almost always smaller and never greater) than that of the vacuum state, directly showing the violation of the Heisenberg uncertainty principle.
Here $\hbar = 1$ and we have taken $\alpha = \sqrt{\ln 2}/2$ (see text). Note that, even for a Gaussian bivariate distribution, the standard deviation, $\sqrt{V(\hat x_\theta)}$, does not, in general, correspond to the distance in the $\theta$ direction from the centroid to the $e^{-1/2}$-contour of the distribution (as plotted in Fig.~\ref{fig:cat_stripes}), though it does for the axes of symmetry of the contour ellipse.}
\label{fig:var_th}
\end{figure}

\section{Comparison with the Wigner Function of the SWV pseudo-state} 
\label{sec:comp}

It was stated by Tsang, below Eq.~(5.11) of Ref.~\cite{Tsa-PRA09}, that, in all cases (not just LGQ systems), $\varrho_{\rm SWD}$ has the same expectation value for $\hat \bx$ as does $\varrho_{\rm SWV}$. That is, integrating the smoothed Wigner distribution times ${\bf x}$ gives the correct values, conditioned on past and future measurement results, for a weak measurement of any components of $\hat \bx$ at time $t$. We verify this by explicit calculation in Appendix~\ref{App_moments}.  Unfortunately, we were misled by this identity to the assumption that $\varrho_{\rm SWD} = \varrho_{\rm SWV}$. Hence, in Refs.~\cite{LCW19,LCW-PRA21}, in all of the expressions for the moments of $\varrho_{\rm SWD}$, we wrongly called these moments of $\varrho_{\rm SWV}$. (Although the first moments are in fact the same, as stated.) Likewise, in plots of what we here call $\digamma_{\rm SWD}(\bx;t)$, we incorrectly identified it as the Wigner function of the Smoothed Weak-Valued pseudo-state. We stress that there were no errors in the calculations of $\rho\sm(t)$, or plots of $W\sm(\bx;t)$, its Wigner function. 

Given their similarity in construction, and the fact that they have the same means in phase-space, it is natural to ask how $\digamma_{\rm SWD}(\bx;t)$ and $W_{\rm SWV}(\bx;t)$ differ. {For the general forms of the characteristic functions of the SWV and SWD states, see Appendix~\ref{app}. While computing $\digamma_{\rm SWD}(\bx;t)$ is simple, computing the Wigner function of the SWV state seems, in general, quite difficult due to the cosine term in \erf{eq:cosine}. As such, we will consider a simple, but telling, example.} The differences between the SWD and the SWV Wigner function will be greatest when $\hat{E}\rfil(t)$ and $\rho\fil(t)$ are pure, as mixture tends to make all smoothed estimates the same~\cite{LCW19}. In this case, arbitrary choices of $\hat{E}\rfil(t)$ and $\rho\fil(t)$ can make the difference between $\digamma_{\rm SWD}(\bx;t)$ and $W_{\rm SWV}(\bx;t)$ drastic. However, we show that a typical filtered state and effect for a LGQ system yield, even when pure, somewhat similar results for  $\digamma_{\rm SWD}(\bx;t)$ and $W_{\rm SWV}(\bx;t)$. But only the former is a Gaussian distribution; the latter has negativities.

The prototypical pure Gaussian state is a coherent state. As such, we will assume that both the filtered state and retrofiltered effect are coherent states. That is, $\rho\fil = \ket{\alpha}\bra{\alpha}$ and $\hat{E}\rfil = \ket{\beta}\bra{\beta}$. Computing the resulting SWV state gives
\beq\label{SWV_coherent}
\varrho\swv = \frac{1}{\cal N} \left(e^{-\beta^*\alpha}\ket{\alpha}\bra{\beta} + e^{-\alpha^*\beta}\ket{\beta}\bra{\alpha}\right)\,,
\eeq
where ${\cal N} = 2{\rm exp}\left[-(|\alpha|^2 + |\beta|^2)/2\right]$. To get some immediate intuition of what this state will look like, we can consider the general \sch~cat state $\rho_{\rm cat} \propto \op{\alpha}{\alpha}+{\red (e^{i\phi}\op{\alpha}{\beta}+ e^{-i\phi}\op{\beta}{\alpha})}+\op{\beta}{\beta}$. We can see that (up to normalization) the SWV state is the interference term (red) of a cat state. In terms of the Wigner function, this term is responsible for the interference fringes that appear between the Gaussian peaks at $\alpha$ and $\beta$. Moreover, these interference fringes are the only part of the cat state where the Wigner function can be negative. Thus, we expect that the Wigner function for the SWV state can contain some negativity (always bearing in mind that one cannot interpret this as Wigner-negativity of a state, since $\varrho\swv$ is a pseudo-state). 
Without loss of generality, we can consider the case where $\beta = -\alpha$, and $\alpha \in {\mathbb R}$. 

For the particular value of $\alpha$, we could take a very large value, and the difference between $W\swv(\bx;t)$ and $\digamma_{\rm SWD}(\bx;t)$ will be staggering. However, the coherent amplitudes of the filtered state and retrofiltered effect are very unlikely to be far apart in reality; in the large $\alpha$ limit, $\rho\fil(t)$ and $\hat{E}\rfil$ would be almost orthogonal, meaning the corresponding measurement record would almost never occur. To consider a more typical value for $\alpha$, let us imagine the set of possible retrofiltered effects as $\{\frac{1}{\pi}\ket{\beta}\bra{\beta}\dd^2\beta\}$. The probability density $\wp(\beta|\alpha)\dd^2\beta$ of a particular effect for a given filtered state $\ket{\alpha}\bra{\alpha}$ is given by 
$\frac{1}{\pi} |\ip{\alpha}{\beta}|^2\dd^2\beta = \frac{1}{\pi} e^{-|\alpha - \beta|^2} \dd^2\beta$.
Evidently, this probability density depends only on $|\alpha - \beta|$, and so we can ask the question: what is the probability that $|\alpha - \beta| < \delta$. Armed with that, we can then ask what is a typical value of $|\alpha - \beta|$,  
in the colloquial sense. Specifically, we ask what is the $\delta$ such that $P(|\alpha - \beta|<\delta) = 0.5$, as half the time we will see differences of this size or smaller. It is easy to show that the answer is $\delta = \sqrt{\ln 2}$. Then, since we are considering $\beta = - \alpha$, we get $\alpha = \sqrt{\ln 2}/2 \approx 0.416$ for a typical situation.

With the particulars dealt with, we can see, in Fig.~\ref{fig:cat_stripes}, the differences between (Left) $W\swv(\bx;t)$  and (Right) $\digamma_{\rm SWD}(\bx;t)$. As expected, we see that both distributions are centered at the same point in phase space (here $q = p = 0$), but the SWV state is clearly non-Gaussian due to the negative regions.
We do reiterate that neither of these distributions correspond to physically valid quantum states. {This can be seen by the fact that both distributions are more condensed around the origin than the vacuum state, as indicated in Fig.~\ref{fig:cat_stripes} by the $e^{-1/2}$ contours. More quantitatively, in Fig.~\ref{fig:var_th}, we see that the quadrature variance $V(\hat{x}_\theta) = \ex{\hat{x}_\theta^2} - \ex{\hat{x}_\theta}^2$, where $\hat{x}_\theta = \hat{q}\cos\theta + \hat{p}\sin\theta$, for both the SWD (blue line) and SWV (red line) state are never greater always than that of the vacuum state (black dashed), and in fact are always smaller (SWD) or almost always smaller (SWV) than it. That is, both cases violate the Heisenberg uncertainty relations.}

\section{Conclusion}
In this paper, we have corrected an error that we made in Ref.~\cite{LCW-PRA21,LCW19} {(and  propagated in \cite{SoroushThesis})} where we mistakenly {\blk equated two distinct methods of semi-classical smoothing}, what we are here calling the smoothed Wigner distribution, with the Wigner distribution of the smoothed weak-valued state. As a result of this mistake, almost all reference to the linear Gaussian smoothed weak-valued state in Ref.~\cite{LCW-PRA21,LCW19} should actually be references to the smoothed Wigner distribution. 
We hope this paper not only corrects any confusion we might have caused by our error, but also draws attention to the remarkable correspondence in the mean noted by Tsang for these two very different generalizations of classical smoothing, a correspondence not shared by any of the other generalizations (defining valid quantum states) discussed in the penultimate paragraph of the introduction~\cite{GueWis15,CGLW21,LASL25}. It also highlights the intricacies and richness that still remains in the field of quantum state estimation using past and future information.

\begin{widetext}
\appendix
\section{Equivalence of the first moments for the SWV state and the SWD}\label{App_moments}
Here, we provide a proof that the SWV state and the SWD share the same first moment, stated by Ref.~\cite{Tsa-PRA09}. For simplicity, we will only prove this for one canonical variable of a single mode system. That is, we will show that
\beq\label{seq:equiv}
\ex{\hat{q}}\swv={\rm Re}\left[\frac{\Tr[\hat{q}\rho\fil\hat{E}\rfil]}{\Tr[\rho\fil\hat{E}\rfil]}\right] = \int\dd q \int\dd p\,\,q\,\, \digamma_{\rm SWD}(q,p) = \ex{q}_{\rm SWD}\,.
\eeq
The extension to arbitrary quadratures of an $n$-mode system is not difficult. 

Let us begin with the left-hand-side of \erf{seq:equiv},
\beq\label{eq:SWVmoment}
{\rm Re}\left[\frac{\Tr[\hat{q}\rho\fil\hat{E}\rfil]}{\Tr[\rho\fil\hat{E}\rfil]}\right] = {\rm Re}\left[\frac{\int\dd q\int\dd p W_{\hat{q}\rho\fil}(q,p) W\rfil(q,p)}{\int\dd q\int\dd p W\fil(q,p) W\rfil(q,p)}\right]\,,
\eeq
where we have used the identity
\beq
\Tr[\hat{A}\hat{B}] = 2\pi\int\dd q\int\dd p \,\, W_{\hat{A}}(q,p) W_{\hat B}(q,p)\,,
\eeq
and $W_{\hat{q}\rho\fil}$ is the Wigner function of the operator $\hat{q}\rho\fil$. It is known \cite{WisMil10} that 
\beq
W_{\hat{q}\rho\fil}(q,p) = \left(q + i\frac{\partial}{\partial p}\right)W\fil(q,p)\,.
\eeq
Noticing that both $W\fil(q,p)$ and $W\rfil(q,p)$ are real, we have 
\begin{align}
{\rm Re}\left[\frac{\Tr[\hat{q}\rho\fil\hat{E}\rfil]}{\Tr[\rho\fil\hat{E}\rfil]}\right] &= \int\dd q\int\dd p \frac{\left(qW\fil(q,p) + {\rm Re}\left[i\frac{\partial}{\partial p}W\fil(q,p)\right]\right)W\rfil(q,p)}{\int\dd q\int\dd p W\fil(q,p) W\rfil(q,p)}\\
& =  \int\dd q\int\dd p\,\,q\,\, \frac{W\fil(q,p)W\rfil(q,p)}{\int\dd q\int\dd p W\fil(q,p) W\rfil(q,p)}\,.
\end{align}
The proof for $\ex{\hat{p}}\swv = \ex{p}_{\rm SWD}$ follows similarly.

\section{Wigner Function of the Smoothed Weak-Valued State}\label{app}
To obtain an expression for the Wigner function, let us begin with the representation of an $n$-mode bosonic quantum state in terms of its characteristic function
\beq
\rho = \frac{1}{(2\pi)^n}\int \dd{\bf k}\,\,\chi({\bf k}) e^{-i{\bf k}\tp\hat{\bx}}\,,\\
\eeq
where the characteristic function is defined as $\chi({\bf k}) = \Tr[\rho e^{i{\bf k}\tp\hat\bx}]$. Thus, for the SWV state, we have
\beq
\begin{split}
\varrho\swv &= \frac{1}{(2\pi)^{2n}(2N)} \int\dd{\bf k}\int\dd{\bf g}  \left[\chi\fil({\bf k})\chi\rfil({\bf g}) + \chi\fil({\bf g})\chi\rfil({\bf k})\right]e^{-i{\bf k}\tp \hat\bx} e^{-i{\bf g}^{\top}\hat\bx}\\
&= \frac{1}{(2\pi)^{2n}(2N)} \int\dd{\bf k}\int\dd{\bf g}  \left[\chi\fil({\bf k})\chi\rfil({\bf g}) + \chi\fil({\bf g})\chi\rfil({\bf k})\right]e^{-i{\bf k}\tp\Sigma{\bf g}/2}e^{-i({\bf k}+{\bf g})\tp\hat\bx}\,,
\end{split}
\eeq
where, in the second line, we have used the Baker-Campbell-Hausdorff identity and $[\hat{\bx}_i,\hat{\bx}_j'] = -i\Sigma_{ij}$. For simplicity, we have taken $\hbar = 1$. Computing the characteristic function for this yields
\beq
\begin{split}
\chi({\bf k}') &= \Tr[\varrho\swv e^{i{\bf k}'^{\top}\hat{\bx}}]\\
& = \frac{1}{(2\pi)^{n}(2N)}\int\dd{\bf k}\left[\chi\fil({\bf k})\chi\rfil({\bf k}'-{\bf k}) + \chi\fil({\bf k}'-{\bf k})\chi\rfil({\bf k})\right]e^{-i{\bf k}\tp\Sigma({\bf k}'-{\bf k})/2}\,,
\end{split}
\eeq
where we have used $\Tr[e^{i{\bf k}\tp\hat{\bx}}e^{-i{\bf g}\tp\hat{\bx}}] = (2\pi)^n\delta({\bf k} - {\bf g})$. This expression can be simplified to 
\beq \label{eq:cosine}
\chi\swv({\bf k}') =\frac{1}{(2\pi)^{n}N}\int\dd{\bf k}\,\,\chi\fil({\bf k})\chi\rfil({\bf k}'-{\bf k})\cos\left({\bf k}\tp\Sigma({\bf k}'-{\bf k})/2\right)\,.
\eeq

We can compare this to the characteristic equation for the SWD, 
\beq
\chi_{\rm SWD}({\bf k}') = \frac{1}{(2\pi)^n N}\int\dd{\bf k} \chi\fil({\bf k})\chi\rfil({\bf k}'-{\bf k})\,.
\eeq 
Clearly, both characteristic equations are quite similar in form with the only difference being the characteristic equation for the SWV state having a cosine modulating the filtered and retrofiltered characteristic equations.


\end{widetext}
%

\end{document}